\documentclass[11pt]{article}

\usepackage[margin=1in]{geometry}
\usepackage{amsmath,amssymb,bm,gensymb}
\usepackage{graphicx}
\usepackage{booktabs}
\usepackage{enumitem}
\usepackage{xcolor}
\usepackage{microtype}
\usepackage[hidelinks]{hyperref}
\usepackage{newfloat}
\DeclareFloatingEnvironment[name=Schematic,fileext=los]{schematic}

\setlist[itemize]{leftmargin=1.4em,itemsep=0.25em,topsep=0.25em}
\setlist[enumerate]{leftmargin=1.7em,itemsep=0.35em,topsep=0.25em}

\newcommand{\lp}{\ell_{\mathrm p}}
\newcommand{\kb}{k_{\mathrm B}}
\newcommand{\kf}{\kappa_{\mathrm f}}
\newcommand{\Rf}{R_{\mathrm F}}
\newcommand{\Kb}{K}
\usepackage{upgreek}
\providecommand{\muup}{\upmu} 
\newcommand{\todo}[1]{\textcolor{blue!70!black}{%
  \ifmmode\text{[TODO: #1]}\else[TODO: #1]\fi}}

\title{\textbf{Bending Rigidity of Smectic Filaments}}

\author{%
Haotian Long$^{1}$, Shamiah A. Williams$^{1,2}$, Chinedum O. Osuji$^{1}$, Christopher A. Browne$^{3,*}$\\[0.5em]
\small $^{1}$Department of Chemical and Biomolecular Engineering, University of Pennsylvania,\\ \small Philadelphia, PA 19104, USA\\
\small $^{2}$Department of Chemical and Petroleum Engineering, University of Pittsburgh,\\ \small Pittsburgh, PA 15260, USA\\
\noindent \small $^{3}$Department of Chemical Engineering, University of Michigan, Ann Arbor, MI 48109, USA
}

\date{}

\begin{document}
\maketitle

\begin{abstract}

Many biological and engineered filaments exhibit internal structure or fluidity that alters their elastic behavior. We study the elasticity of microscale filamentous smectic condensates, grown during liquid-liquid crystal phase separation. We use a Fourier analysis of their thermal bend fluctuations to measure their persistence length, which follows a non-classic linear scaling with their cross-sectional area, in contrast to the classic quadratic scaling for elastic filaments. We rationalize this scaling using the Helfrich energy to compute the free energy of bending a filament with internal smectic fluid ordering. Our analysis then provides a direct method to measure the bend modulus $K$, which quantifies the rigidity of smectic layers; providing the first method to directly measure $K$ for dispersed smectic phases. We quantify the softening of filaments in the presence of shorter alkyl-tail mesogen dopants, decreasing $K$ by up to $\sim35\%$. Out results suggest how nanoscale smectic organization and molecular composition determine filament-scale elasticity, informing the control of filament mechanics and self-assembly in biomimetic and engineered materials.
\end{abstract}

\noindent \small ${}^{*}$ To whom correspondence should be addressed. Email: cabrowne@umich.edu

\section*{Introduction}
Slender filaments and fibers are ubiquitous structural elements in biological, synthetic, and self-assembled materials. The mechanical properties of these filaments play key roles in their structure, assembly, and function. For example, cytoskeletal networks are comprised of elastic actin \cite{gittes1993flexural,ott1993measurement,isambert1995flexibility}, microtubule \cite{venier1994analysis,pampaloni2006thermal,wisanpitayakorn2022measurement}, and intermediate filaments \cite{guzman2006exploring,smoler2020apparent,block2018viscoelastic} which form interpenetrating viscoelastic networks \cite{lin2010divalent,shen2021effects}. Condensates formed by liquid-liquid phase separation can tune the rigidity of these cytoskeletal networks \cite{basu2025vimentin}, and mediate the assembly of filamentous cellular junctions  \cite{pombo2024membrane}. At larger scales, biological fibers and silks exhibit changing elastic behavior during drawing \cite{vollrath2001liquid,baer2017mechanoresponsive,baer2023internal}, with implications for fabrication biomimetic materials \cite{rising2023biological,harrington2024fluid}.

Filament flexibility is quantified by the bending (or flexural) rigidity $\kf$, defined through the elastic energy per unit length $\tfrac{1}{2}\kf C^{2}$ stored when the filament is bent to a curvature $C$. For microscale filaments, thermal energy is often sufficiently large to induce bend fluctuations on a characteristic length scale $\lp=\kf/(\kb T)$, termed the persistence length \cite{gittes1993flexural,Brangwynne2007}.  Classically, a homogeneous elastic solid filament with an isotropic Young's modulus $E$ has a bending rigidity and persistence length scale that scale with the cross-sectional area $\sim E\,A^2$ and radius $\sim E\,\Rf^4$. The steepness of this scaling arises from the requisite material compression and expansion in the inner and outer portions of a bend, respectively.

This classic scaling can be violated in systems with internal structure. For example, membrane tubes exhibit a bending rigidity that scales linearly with radius $\sim\Rf$~\cite{Derenyi2002,Yamamoto2012,Barooji2016}, proportional to the annular cross-sectional area of the single lipid membrane resisting deformation, and hence typically exhibit very short persistence lengths on the order of micrometers. Myelin sheaths, comprised of many lipid membranes, behave closer to composite elastic structures due to inter-lamellar binding by Myelin Basic Protein (MBP) \cite{mueller1999force}, though loss or modification of MBP in disease pathologies allows layer slippage, leading to softening of the myelin composite structure \cite{weil2016loss,mckie2023altered}. For reconstituted myelin figures and analogous lipid smectics, the resulting structures are understood to exhibit an elastic rigidity while maintaining internal fluidity \cite{huang2005confined,Zou2006,huang2006theory,reissig2010three}. 

Similar microscale filaments with non-classical elasticity can form in a variety of liquid crystalline materials, including phase-separated nematics \cite{wei2019molecular,peddireddy2021self}, core-shell liquid crystalline fibers \cite{buyuktanir2010self,wang2016morphology,sharma2018electrospun}, re-entrant smectic phases \cite{weinan1999dynamics,arora1989reentrant,pratibha1992cylindrical}, lyotropic phase-separated smectics \cite{peddireddy2013myelin,peddireddy2021self,nguindjel2024reconfigurable,winkens2022self,singh2025predator}, and thermotropic phase separating smectics \cite{naito1995preferred,naito1997pattern,todorokihara2004periodic,Jakli2003,jakli2000helical,Morimitsu2024,browne2026arrested,browne2025structural,peddireddy2013myelin}. These materials have potential uses in assembling biomimetic fibers and silks, which similarly undergo liquid-liquid crystal phase separation during drawing \cite{rising2023biological,harrington2024fluid}. Liquid crystal filaments also have desirable optical properties \cite{humar2016liquid,muvsevivc2016liquid}, with potential uses for structural color \cite{park2019directed,zhang2014helical}, optical resonators \cite{humar2009electrically}, lasers \cite{papivc2021topological,zheng2015wide,humar20103d}, and light guides \cite{peddireddy2013lasing,papivc2021topological,takenaka2020nematic,d2021light}. These materials are particularly appealing because of their ability to self-assemble into hierarchical structures including dendrites \cite{wei2019molecular,peddireddy2021self} and networks \cite{Morimitsu2024,browne2025structural,browne2026arrested}. However, predictive control of these structures remains limited, in large part because quantitative comparison of elasticity between compositionally-similar systems remains challenging. The elasticity of these materials, quantified by the bend modulus $K$, is traditionally measured from the line tension of dislocation lines or from the Fr\'{e}edericksz transition under an applied magnetic field \cite{Oswald2014, Zywocinski2000}; both of which require specialized sample preparation and prior knowledge of other material parameters, like the magnetic polarizability, which can also be challenging to measure.

Here, we directly measure the bend modulus $K$ for smectic filaments from their thermal bend fluctuations. We grow filaments using liquid-liquid crystal phase separation and extract the thermal persistence length $\lp$ using a Fourier analysis to separate thermal bending fluctuations from their static contour. We observe a non-classic elastic scaling of $\lp\sim\Rf^2$, rationalized by an analysis of the Helfrich energy of a bending smectic filament. This procedure enables direct facile measurement of the smectic bend modulus $K$ without prior knowledge of any other material parameters. We use this method to measure the variation of bend modulus $K$ with the addition of dopants, rationalizing differences in bulk phase separation behavior and self assembly of networked architectures. These results suggest methods to quantitatively tune the elasticity and self-assembled structures of smectic materials.

\section*{Results}

\textbf{Bending rigidity of smectic filaments exhibits anomalous scaling.} We measure the elasticity of smectic filaments from their thermal persistence length. We grow filaments directly from the isotropic liquid of mass fraction $\omega=0.30$ 12OCB (4$'$-cyano-4-$n$-dodecyloxybiphenyl, TCI Chemicals) and balance squalane (TCI Chemicals).  Samples are cooled using a temperature control stage (Linkam THMS 600) mounted on a polarized optical microscope (Zeiss Axio Observer.A1) with a 20$\times$ objective for visualization. Videos are recorded using a CMOS camera (Thorlabs Kiralux CS505CU) capturing 10 frames per second and a field of view 350$\times$350 µm. Upon cooling, this binary system undergoes liquid-liquid crystal phase separation, forming a dispersed 12OCB-rich phase ($\omega_\text{SmA}=0.80$) with smectic A liquid crystal ordering and a continuous 12OCB-poor ($\omega_\text{I}=0.38$) phase \cite{browne2025structural}. As previously characterized, the dispersed phase grow as slender filaments, driven by the isotropic to smectic A ordering which occurs concomitant with the phase separation \cite{Morimitsu2024}. Filaments nucleate with a distribution of radii $R_\text{F}\sim0.5$ to $4~\mathrm{\mu m}$, but once formed elongate with constant radius \cite{browne2026arrested}. We cool the system just below binodal, resulting in a large number of sparse, straight filaments with lengths $L\lesssim 100~\mathrm{\mu m}$ insufficient to hydrodynamically buckle \cite{Morimitsu2024}. We then hold the temperature constant at $T\approx63\degree\text{C}$, allowing filaments to equilibrate their growth. Filaments remain sufficiently sparse that they do not coalesce \cite{browne2026arrested} over the timescale of the experiment. 

Our filaments show visible thermal bending fluctuations on the timescale of seconds (Fig.~\ref{fig:master}a, supplementary movie 1). We use these thermal fluctuations of filaments to characterize their persistence length. We use intensity-envelope detection (SI Section S1.3) to extract filament centerline contours $s(t)$ for a duration of about $5~\text{min}$, and decompose the fluctuations into Fourier-modes following established methods for cytoskeletal filaments \cite{gittes1993flexural,Brangwynne2007}. We compute the tangent angle $\theta(s,t)$ along the contour and decompose it into Fourier cosine modes:

\begin{figure}[!t]
\centering
\IfFileExists{Fig1.pdf}{\includegraphics[width=6.5in]{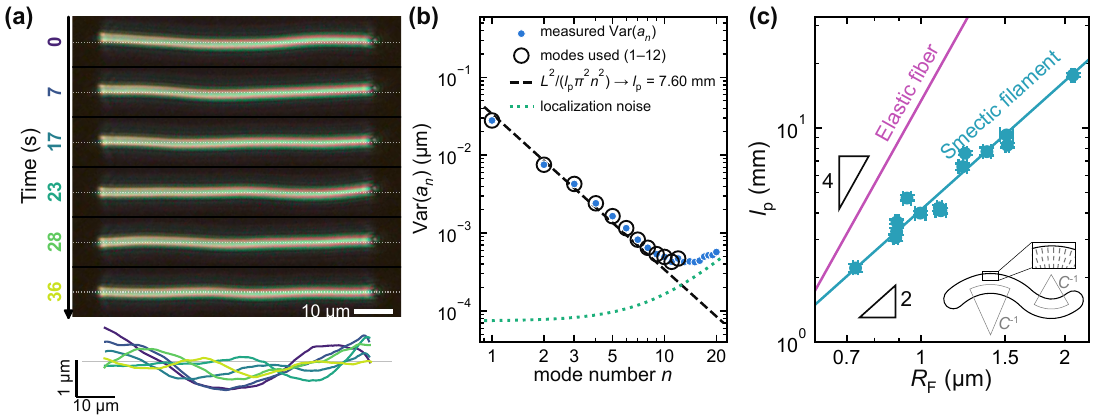}}{\framebox[\textwidth]{\rule{0pt}{8cm}}}
\caption{\textbf{Smectic filament persistence length measured from thermal fluctuations.}
(a) Polarized optical micrographs of a representative filament ($\Rf = 1.24~\muup$m, $L = 66 ~\muup$m) show thermal bend fluctuations on the timescale of seconds. The dotted line is a fixed reference. Scale bar, 10~$\muup$m, crossed polarizers at approximately 45\degree-135\degree Bottom: transverse deviation $\Delta y$ of the tracked centerline from its time-averaged contour along the arc length $s$ for the same six frames displayed. Vertical displacement is exaggerated for visibility.
(b) Variance spectrum $\operatorname{Var}(a_n)$ versus mode number $n$ for the filament in (a). Open circles mark the modes used in the fit ($n = 1$--12), over which the thermal and localization-noise contributions were fitted jointly (SI). The dashed line is the thermal contribution $L^{2}/(\lp\pi^{2}n^{2})$ of eqn~\eqref{eq:equipartition}, which gives $\lp = 7.60$~mm (95\% CI 7.46--7.77~mm); the dotted line is the localization-noise contribution, which dominates beyond $n \approx 12$. 
(c) Persistence length $\lp$ for many filaments of varying radii $\Rf$ (vertical error bars, 95\% bootstrap CI of $\lp$; horizontal error bars, frame-to-frame standard deviation of $\Rf$). All filaments are dopant-free pristine 12OCB. Solid lines show $\Rf^4$ (pink) and $\Rf^2$ (blue) scaling expected for an elastic solid and our prediction for a smectic filament (eqn~\eqref{eq:smectic}), respectively. The free exponent fitting gives an exponent of $1.91 \pm 0.13$, in good agreement with $\Rf^2$. Fit of eqn~\eqref{eq:smectic} gives $K = 48.9 \pm 1.6~\mathrm{pN}$.}
\label{fig:master}
\end{figure}

\begin{equation}
\theta(s,t)=\sqrt{\frac{2}{L}}\sum_{n\ge1}a_n(t)\cos\!\left(\frac{n\pi s(t)}{L}\right),
\label{eq:modes}
\end{equation}
\noindent where the mode amplitude $a_n(t)$ is further decomposed into the time-averaged $a_n^{0}=\langle a_n\rangle_t$ and fluctuating $\delta a_n(t)=a_n(t)-a_n^{0}$ components, where $\langle\cdot\rangle_t$ indicates an average taken over time. Tracking each filament over time lets this decomposition isolate the thermal fluctuations $\delta a_n(t)$ from the static bends $a_n^{0}$ frozen in during growth, which bias persistence lengths from static shape measures \cite{ott1993measurement,wisanpitayakorn2022measurement} toward low values \cite{Schellman1995,Bednar1995}, especially for stiff filaments ($L \ll \lp$) here. Equipartition theorem gives \cite{gittes1993flexural,Brangwynne2007}: 

\begin{equation}
\operatorname{Var}(a_n)\equiv\langle\delta a_n^{2}\rangle_t=\frac{\kb T}{\kf}\left(\frac{L}{n\pi}\right)^{2}=\frac{1}{\lp}\left(\frac{L}{n\pi}\right)^{2},
\label{eq:equipartition}
\end{equation}
\noindent where the right hand side uses the definition of persistence length $\lp\equiv\kappa_f/k_\text{B}T$. 
Equation~\eqref{eq:equipartition} holds only for purely thermal bending, for which the low-mode fluctuations follow an initial $\langle\delta a_n^{2}\rangle_t\sim n^{-2}$ scaling (Fig.~\ref{fig:master}b). We therefore use this scaling as an acceptance test and keep only filaments whose spectra pass it (SI Section S3.3). At higher modes the spectrum departs from this scaling because of localization noise \cite{gittes1993flexural}, which we fit together with $\lp$ (SI Section S3.2). Our fitting procedure, acceptance criteria and the sensitivity of $K$ to the noise treatment are described in detail in SI Sections S3.2--S3.4.

As expected, filaments exhibit a bending rigidity and persistence length which grows with radius $\Rf$
(Fig. \ref{fig:master}c). Unlike solid elastic cylindrical fibers, which universally exhibit $\lp\sim R_\text{F}^4$ \cite{LandauLifshitz1986,Howard2001}, these cylindrical smectic A filaments exhibit a significantly softer scaling $\lp \sim\Rf^{\alpha}$ of $\alpha = 1.91$ (95\% CI 1.69--2.07). Thus, smectic A filaments exhibit an apparent elasticity, but with a non-classic scaling that deviates from solid elastic fibers.

\textbf{Modeling bending rigidity of smectic filaments}. 
The classic $\lp\sim R_\text{F}^4$ scaling arises for filaments of compositionally-homogeneous isotropic elastic
material. In this case, the bending rigidity $\kf=EI$ is given by the product of the Young's modulus $E$ and the second moment of
area of the cross-section, $I=\int y^{2}\,\mathrm{d}A$, where $y$ is the distance from the neutral
axis, $I=\pi\Rf^{4}/4$ for a cylinder of radius $\Rf$~\cite{gittes1993flexural, LandauLifshitz1986}. Hence, solid elastic filaments exhibit a bending rigidity and persistence length that scales with $\Rf^4$, or equivalently $A^2$: $\lp \;=\; (\pi/4)(\kb T)E\,\Rf^{4}$. 

We rationalize the anomalous scaling by computing the free energy associated with these bending modes distorting the underlying smectic order. The elastic penalty of smectic layer deformations is captured by the Helfrich free energy \cite{Helfrich1973}:
\begin{equation}
F_\text{H} = \frac{\Kb}{2}\int H^{2}\,\mathrm{d}V,
\qquad
H = \tfrac{1}{2}\left(c_{1}+c_{2}\right),
\label{eq:curvenergy}
\end{equation}
where $H$ is the mean curvature of the layers, with principal curvatures $c_{1}$ and $c_{2}$, and $K$ is the bend modulus which quantifies the penalty of bending smectic layers. This expression is a simplification of the Frank-Oseen energy for liquid crystals, which represents this as a splay modulus $K_{1}\equiv\Kb/4$ which quantifies the penalty of splaying the smectic director field $\mathbf{n}$ as equivalent energy $\tfrac{1}{2}K_{1}(\nabla\cdot\mathbf{n})^{2}$ since $\nabla\cdot\mathbf{n} = 2H$. The director bend and director twist moduli of smectics are typically sufficiently large that these director deformation modes are negligible. Filaments exhibit homeotropic anchoring, enforcing a curvature $H=1/(2r)$ which varies in the radial dimension $r$ \cite{Morimitsu2024}. Smectic order is lost within a defect core running axially along the filament centerline with radius $R_\text{C} \lesssim d $, which is expected to be smaller than the smectic layer spacing $d=4.9~\text{nm}$, measured by small angle x-ray scattering (SAXS) \cite{browne2025structural}, and hence $R_\text{F}/R_\text{C}\ll1$. Integrating for a perfectly straight smectic filament gives a ground state energy $F_\text{H}(0)=(1/4)(\pi K L)\ln(R_{\text{F}}/R_\text{C})$.

To obtain an expression for the bending rigidity, we compute the energy $F_\text{H}(C)$ to bend the smectic filament with a uniform constant curvature $C$ along its entire centerline. Each layer at radius $r\in[R_\text{c},R_\text{F}]$ forms a toroidal section with minor radius $r$ and major radius $C^{-1}$. Integration of the resulting curvature $H(r,\theta)$, which additionally varies in the azimuthal direction $\theta$, using Eq.~\eqref{eq:curvenergy} (SI) gives: 
\begin{equation}
\Delta F_{\mathrm{bend}}(C) = F_\text{H}(C) - F_\text{H}(0)
\approx\frac{\pi}{16}\,\Kb\,L\,\Rf^{2}\,C^{2},
\label{eq:Fbent}
\end{equation}

\noindent which has been simplified in the limit $R_\text{C}/R_\text{F}\lesssim5\cdot10^{-3}\ll1$ (SI Section S2.2). The quadratic dependence on $C$ matches the expectation that the energy penalty must be symmetric and positive for bends of opposite-signed curvature. The quadratic dependence on $C$ further matches the energy for worm-like chains, which have $\Delta F_{\mathrm{bend}}(C)=\tfrac{1}{2}\kf\,L\,C^{2}$ \cite{gittes1993flexural,Saito1967}. Thus, smectic filaments have an elastic behavior similar to worm-like chains with a comparable bending rigidity of $k_f=(\pi/8)\Kb\Rf^{2}$, and a persistence length of:

\begin{equation}
\lp = \frac{\pi}{8}\frac{\Kb}{\kb T}\Rf^{2},
\label{eq:smectic}
\end{equation}
\noindent consistent with our experimentally observed scaling (Fig. \ref{fig:master}c).

This analysis further indicates that thermal measurements of the persistence length enable direct measurement of the smectic bend modulus $K$ of the smectic phase, without knowledge of any other material parameters. Fitting Eq.~\eqref{eq:smectic} to the data in Fig.~\ref{fig:master}c gives $K = 48.9 \pm 1.6~\mathrm{pN}$ (best fit $\pm$ bootstrap standard error; SI) for the 12OCB-Squalane smectic A phase (12OCB content $\omega_\text{SmA}\approx0.80$ balance Squalane measured by SAXS \cite{browne2025structural}). This measure is of consistent order with estimates of the corresponding director splay constant, $K_{1} = \Kb/4 \sim 10$~pN for similar bulk cyanobiphenyls in the smectic-A phase and near the nematic--smectic-A transition \cite{Oswald2014, Zywocinski2000}. Thus, our analysis of thermal filament fluctuations provides a straightforward and precise method to directly measure $K$ for smectics that undergo LLCPS, with no need for external calibrations.

\textbf{Mesogen dopants tune bend modulus.} Our analysis indicates that the bending rigidity and persistence length of smectic filaments emerges directly from the bend modulus $K$ of smectic layers. We expect changes in the smectic layer thickness and composition to tune this bend modulus. Indeed, previous reports suggest that varying solvent affinity alters swelling of the smectic layers and consequently the rigidity of condensate structures \cite{browne2025structural}. 

\begin{figure}[!t]
\centering
\IfFileExists{Fig2.pdf}{\includegraphics[width=\textwidth]{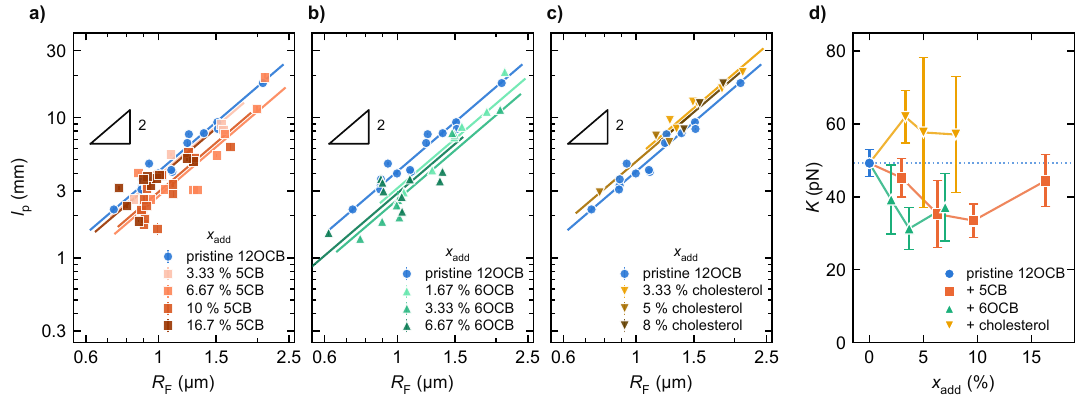}}{\framebox[\textwidth]{\rule{0pt}{5cm}}}
\caption{\textbf{Composition tunes bend modulus.}
Persistence length for 12OCB filaments of various radii doped with (a) 5CB (squares; $x_{\mathrm{add}} = 3.33$--16.7\%), (b) 6OCB (triangles; $x_{\mathrm{add}} = 1.67$--6.67\%) and (c) cholesterol (inverted triangles; $x_{\mathrm{add}} = 3.33$--8\%). Symbol shade darkens with $x_{\mathrm{add}}$. Data for pristine 12OCB (blue circles) duplicated from Figure~\ref{fig:master}c in each panel for reference. Lines are best fit to equation~\eqref{eq:smectic} for each composition with fixed $R_\text{F}^2$ power and $K$ as the sole fitting parameter.
(d) Bend modulus $\Kb$ as a function of added dopant percent $x_{\mathrm{add}}$ for 5CB (orange squares), 6OCB (green triangles) and cholesterol (yellow inverted triangles): mean of the per-filament values of $\Kb$ from eqn~\eqref{eq:smectic}. Error bars show the 95\% confidence interval of the mean. The dotted line marks the value for pure 12OCB, and the connecting lines are guides to the eye.}
\label{fig:composition}
\end{figure}

We directly test this hypothesis by doping the smectic with two shorter-tailed mesogens 5CB (4$'$-cyano-4-$n$-pentylbiphenyl, TCI Chemicals) and 6OCB (4$'$-cyano-4-$n$-hexyloxybiphenyl, TCI Chemicals), fixing total mesogen loading ($\omega_{\mathrm{M}} = 30$~wt\%), and growing filaments of varying radii and compute their persistence length as before. Across all samples, the persistence length and bending rigidity measurements are consistent with our predicted $\sim R_\text{F}^2$ scaling, with a common exponent $\alpha = 1.88 \pm 0.10$ (Fig. \ref{fig:composition}A--C), enabling direct measurement of $K$ as a function of dopant addition. For both nematic dopants, $K$ decreases by up to $\sim35\%$ for $x_\text{add} \approx 3$--10\%, and partially recovers at the highest 5CB loading ($x_\text{add} = 16.7\%$; Fig. \ref{fig:composition}D). At sufficiently high dopant loadings, the dopant alters the liquid-liquid crystal phase separation sufficiently that filaments no longer form, precluding this method from measuring $K$.

We similarly measure changes in $K$ with doping of cholesterol (Sigma Aldrich), which is known to exhibit a robust stiffening effect on a wide range of lipid membranes \cite{chakraborty2020cholesterol,kumarage2025cholesterol,doole2022cholesterol}. As expected, cholesterol stiffens the filaments without altering the $\sim R_\text{F}^2$ scaling (Fig.~\ref{fig:composition}C), with $K = 57$--$62~\mathrm{pN}$ at $x_\text{add} = 3.33$--8\%, about 20\% above pristine 12OCB (Fig.~\ref{fig:composition}D). The broad bidirectional tunability of $K$ with dopants indicates a new avenue for quantitative control of filament rigidity.

\section*{Discussion}

We directly measure the persistence length of smectic filaments, grown by thermotropic liquid-liquid crystal phase separation, by Fourier analysis of their thermal bend fluctuations. This analysis reveals that smectic filaments exhibit a non-classic elastic scaling of $\lp\sim\Rf^2$, in contrast to the $\sim\Rf^4$ scaling for elastic solids. We rationalize this scaling using the Helfrich energy of a bending smectic filament. Approximations in this analysis are valid for smectic filaments with a small defect core and under weak bending (SI Section S2.2), both of which are well justified in our experiments, for which $R_\text{C}/\Rf \sim 10^{-4}$ and $C\Rf \sim 10^{-2}$. The softer scaling observed is ultimately enabled by the fluidity of the smectic within layers. Flow of mesogen molecules within layers relieves compressive strain in the inner and outer portions of bends. Thus, in contrast to elastic solids, smectic filaments have no strain cost to bending, and are only penalized by layer bend. Integration of this Helfrich layer bending energy over layers quantitatively predicts the observed $\lp\sim\Rf^2$ scaling. We expect a similar scaling to arise in mylein figures \cite{huang2005confined,Zou2006,huang2006theory,reissig2010three} and in myelin sheaths in the absence of myelin binding protein (MBP) \cite{weil2016loss,mckie2023altered}, since these systems exhibit an analogous inter-layer fluidity and bending energy that depends only on layer bend and not compression. However, to our knowledge, this scaling has not been measured experimentally.

Our analysis provides a method to directly measure the bend modulus $K$ from these persistence length measurements. To our knowledge, this is the only existing method to directly measure $K$ for a dispersed smectic phase. This method has some advantages over existing methods utilizing the Fr\'{e}edericksz transition, which require prior knowledge of the magnetic susceptibility and careful control of substrate surface anchoring \cite{Oswald2014}. We use this method to directly measure the effect of mesogen dopants on filament rigidity. In the presence of $\sim3$ to $10\%$ of shorter-tailed nematic dopants, we observe up to a $\sim35\%$ decrease in $K$. The similarity in response to different shorter-tailed mesogens may be linked to a shared mechanism of softening. Both short-tailed mesogens are nematic-forming in isolation, and as such have a significantly lower penalty to splay deformations, with $K_1\equiv K/4$ $\sim 100\times$ lower for nematics. Additionally, the mismatch in mesogen alkyl tail size may effectively lower the local mesogen density, enabling localization of layer bend in the vicinity of shorter tails. Lastly, the presence of dopants may alter the mutual solubility of the solvent squalane within the smectic phase, which is known to alter the layer spacing and potentially contribute to softening $K$ \cite{browne2025structural}. In contrast, cholesterol is observed to stiffen these filaments by $\approx20\%$, analogous to its stiffening effect on a wide range of lipid membranes \cite{chakraborty2020cholesterol}. The stiffening effect of cholesterol is likely linked to its tendency to increase local packing density, penalizing bend modes \cite{kumarage2025cholesterol,doole2022cholesterol}. We anticipate that tunability of bend modulus can be used to control the morphology and dynamics of phase separating smectics. Bend modulus plays a known direct role in mediation the formation of helical coils \cite{browne2026arrested}, and is expected to play a broader mediating role in the formation of other morphologies \cite{browne2025structural} and the architecture of condensate networks \cite{Morimitsu2024}. 

These results suggest more generally that internal structure, such as liquid crystallinity, can alter the bulk elastic behavior of microscale filaments. These differences in elasticity, and rigidity scaling may have important consequences in other filamentous biomaterials: for example, in the loss of inter-layer binding in myelin disease pathologies \cite{weil2016loss,mckie2023altered}; during the transient formation of biological \cite{vollrath2001liquid,baer2017mechanoresponsive,baer2023internal} or biomimetic \cite{rising2023biological,harrington2024fluid} drawn fibers; and in biofilaments that couple to other condensate phases \cite{basu2025vimentin,pombo2024membrane}. Understanding and controlling filament dynamics and self-assembly may enable to design of complex architectures \cite{Morimitsu2024} with interesting optical properties for photonic materials \cite{humar2016liquid,muvsevivc2016liquid}.

\section*{Acknowledgments}. 

\noindent\textbf{Funding}. The authors acknowledge support from the National Science Foundation through DMR-2309043. Research reported in this publication was supported by the National Institutes of Health under Award Number U54CA280805. The content is solely the responsibility of the authors and does not necessarily represent the official views of the National Institutes of Health.

\noindent\textbf{Author contributions.} {H.L. and C.A.B.. designed the experiments; H.L. and S.A.W. performed all experiments and analyzed all data; H.L. developed new code; H.L., C.O.O., and C.A.B. designed the theoretical models, discussed the results and implications, and wrote the manuscript; C.O.O. and C.A.B. designed and supervised the overall project.}

\noindent\textbf{Author ORCIDs.}
Haotian Long 0009-0003-2528-3131; 
Shamiah A. Williams 0009-0009-7202-9930;
Chinedum O. Osuji 0000-0003-0261-3065;
Christopher A. Browne 0000-0002-3945-9906;\\

\noindent\textbf{Conflicts of interest.} There are no conflicts to declare.

\noindent\textbf{AI usage.} During the preparation of this manuscript, the authors used Claude (Anthropic) to assist in developing and debugging custom scripts for data processing, analysis, and visualization; and during editing to improve the clarity, grammar, and readability of the text. All final editing of text was done by the authors without AI assistance. All analytical procedures were designed by the authors. All AI-assisted code was reviewed and validated by the authors. The authors take full responsibility for the accuracy of the analyses, the scientific interpretations, and the conclusions.

\noindent\textbf{Data availability}. All study data are included in the article and/or supporting information. Code used to extract contours and persistence length, the tracked centerlines of all analyzed filaments, and sample movies are posted at the online repository \url{https://github.com/Haotian77/smectic-filament-bending}. References 72–73 are cited in the Supplementary Information (SI).

\bibliographystyle{unsrt}
\bibliography{refs}


\newpage

\renewcommand\thefigure{S\arabic{figure}}
\renewcommand\theequation{S\arabic{equation}}
\renewcommand\thesection{S\arabic{section}}
\renewcommand\thesubsection{S\arabic{subsection}}

\setcounter{secnumdepth}{4}
\setcounter{page}{1}
\setcounter{figure}{0}
\setcounter{section}{0}
\setcounter{table}{0}

\begin{center}
{\Large\bfseries Supplementary Information}\\[0.6em]
{\large Radius Scaling of Bending Rigidity in Smectic Filaments}\\[0.6em]
Haotian Long, Shamiah A. Williams, Chinedum O. Osuji, Christopher A. Browne
\end{center}





\section{Materials and methods}
\label{si:methods}

\subsection{Materials and sample preparation}
The mesogenic liquid crystal forming the filaments is 4$'$-cyano-4-$n$-dodecyloxybiphenyl (12OCB, TCI Chemicals). The isotropic solvent squalane (TCI Chemicals), the additive dopants 4$'$-cyano-4-$n$-pentylbiphenyl (5CB, TCI Chemicals) and 4$'$-cyano-4-$n$-hexyloxybiphenyl (6OCB, TCI Chemicals), and cholesterol (Sigma Aldrich) were used as received. For samples containing an additive, its fraction within the liquid-crystal component was defined as $x_{\mathrm{add}} = m_{\mathrm{add}}/(m_{\mathrm{12OCB}} + m_{\mathrm{add}})$, and the total liquid-crystal mass fraction as $\omega_{\mathrm{LC}} = (m_{\mathrm{12OCB}} + m_{\mathrm{add}})/(m_{\mathrm{12OCB}} + m_{\mathrm{add}} + m_{\mathrm{solvent}})$. All mixtures were prepared at $\omega_{\mathrm{LC}} = 30$~wt\%, with 5CB at $x_{\mathrm{add}} = 3.33$, 6.67, 10 and 16.7\% 6OCB at $x_{\mathrm{add}} = 1.67$, 3.33 and 6.67\% and cholesterol at $x_{\mathrm{add}} = 3.33$, 5 and 8\%. Each mixture was heated to 100~$^\circ$C, above its clearing temperature, and mixed for 5~min to produce a homogeneous isotropic phase. The homogeneous mixture was loaded into a Hele-Shaw cell consisting of two cleaned microscope coverslips with a spacing of 20~$\mu$m. Smectic filaments were formed by cooling the sample from 100~$^\circ$C into the smectic--isotropic coexistence region at 1~$^\circ$C~min$^{-1}$. After reaching the imaging temperature, the sample was held for 5~min to reach a steady state before movies were acquired.

\subsection{Optical microscopy and movie acquisition}
Filaments were imaged using a polarized optical microscope (Zeiss) equipped with 5$\times$ and 20$\times$ objectives and a temperature-controlled stage (Linkam Scientific THMS600). The calibrated image scale was 0.17~$\mu$m~pixel$^{-1}$ with 20X objective. Movies were mostly recorded at 10~frames~s$^{-1}$ with an exposure time of 10~ms, for about 5~min (analysed duration 1.3--6.5~min, median 4.8~min), to accumulate statistics for each bending mode. Filaments showing measurable retraction, directed motion, coiling, or strong buckling during the acquisition were excluded. Temperature drift during a movie was less than 0.1~K.

\subsection{Filament tracking and persistence length extraction}
\begin{figure}[!tbp]
\centering
\includegraphics[width=\textwidth]{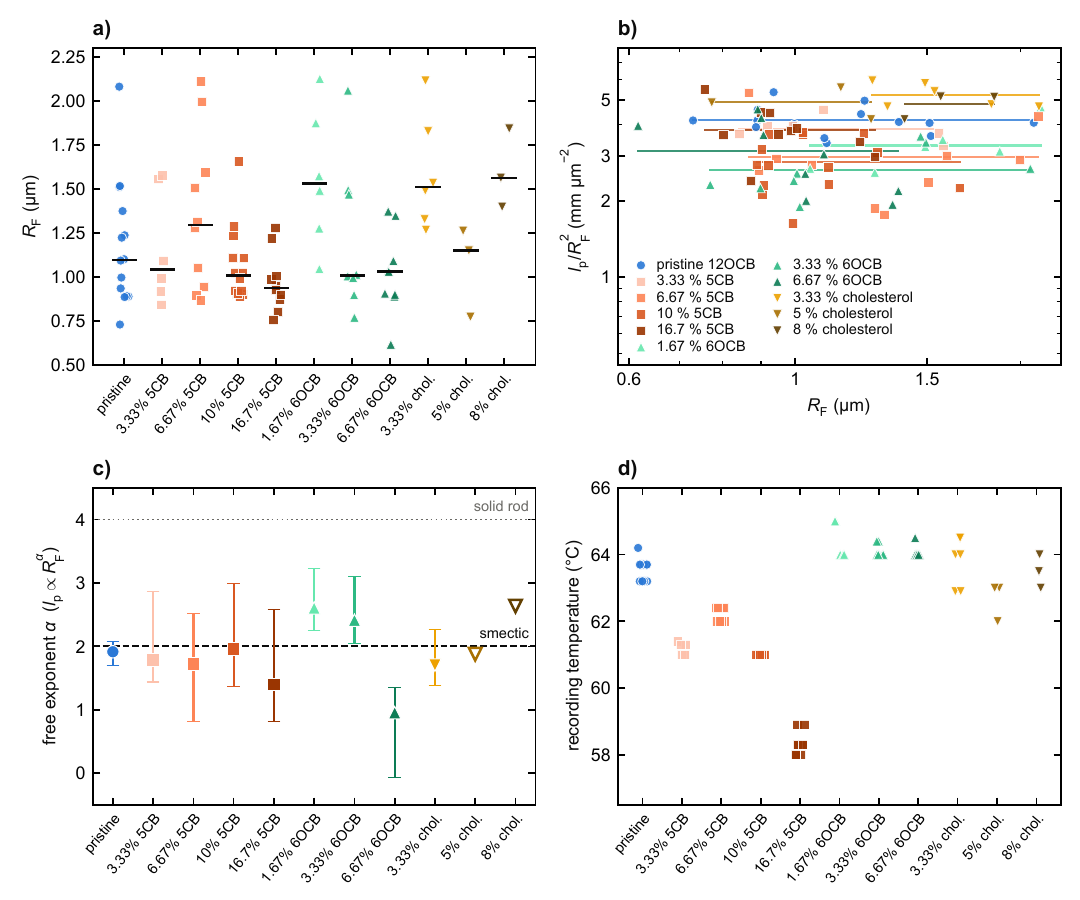}
\caption{\textbf{Sample checks.} (a) Radius $\Rf$ of each accepted filament; bars, sample median. (b) Compensated persistence length $\lp/\Rf^{2}$ versus $\Rf$; lines, sample mean over the radius range of the sample. (c) Free exponent $\alpha$ of $\lp \propto \Rf^{\alpha}$ with its 95\,\% confidence interval from 4000 bootstrap resamples of filaments; open symbols, samples with fewer than five filaments or a radius range below a factor of 1.5, shown without an interval. Dashed line, $\alpha = 2$ (smectic filament, eqn~(5) of the main text); dotted line, $\alpha = 4$ (solid rod). (d) Recording temperature. All $\lp$ from the reported fit (Section~S3.2); chol., cholesterol.}
\label{fig:si-checks}
\end{figure}

\begin{figure}[!tbp]
\centering
\includegraphics[width=\textwidth]{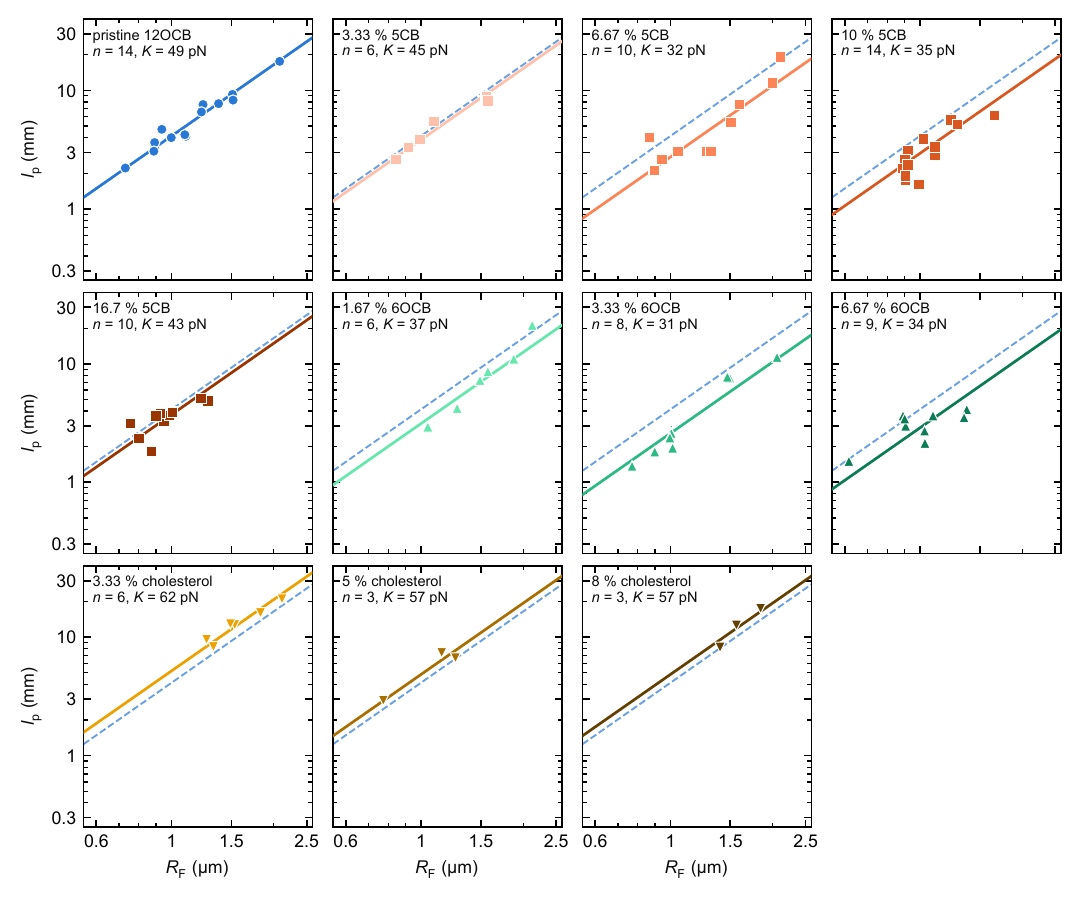}
\caption{\textbf{Persistence length versus radius for each sample.} Symbols, $\lp$ of each accepted filament from the reported fit, with its 95\,\% confidence interval. Solid lines, $\lp = \pi \Kb \Rf^{2}/8\kb T$ with $\Kb$ from the weighted prefactor of the sample (value in the panel); dashed lines, the same for pristine 12OCB.}
\label{fig:si-lpR}
\end{figure}

In every frame, the filament centerline was extracted by intensity-envelope detection and parameterized by arc length $s$, and the filament radius $\Rf$ was taken as half the median full width at half maximum of the transverse intensity profile along the filament over all frames. Because some filaments slowly lengthen or shorten during acquisition, all frames were truncated to a common analysis length $L$, so that every mode amplitude is computed for the same material segment. The tangent angle was decomposed into the cosine modes of eqn~(1) of the main text following Gittes \emph{et al.}\cite{gittes1993flexural}, with each contour resampled into $N$ segments (spacing 0.25~$\mu$m, so that $N = L/0.25~\mu\mathrm{m} \approx 60$--400 for $L = 15$--100~$\mu$m; modes up to $n = 20$ were computed), and the variance of each amplitude was computed about its time average. Each spectrum was first screened for the widest window of modes satisfying pre-declared acceptance criteria (thermal slope, plateau consistency and mode independence; Section~S3.3). For accepted filaments, the reported $\lp$ comes from a joint fit of eqn~(2) of the main text plus a localization-noise term over the modes from the first mode of that window up to $n = 12$ (Section~S3.2), with $\lp$ and an effective noise level $\sigma$ as the only free parameters; the fitted $\sigma$ was well below the pixel size of 0.17~$\mu$m (median 36~nm, interquartile range 25--49~nm, 90th percentile 67~nm, maximum 217~nm). Uncertainties on $\lp$ are 95\% confidence intervals from bootstrap resampling of frames. Of 968 candidate filaments detected and tracked in 123 movies (67 in pure 12OCB, 429 in the 5CB series, 212 in the 6OCB series and 260 in the cholesterol series), 89 passed the acceptance criteria (14 pure 12OCB, 40 5CB-doped, 23 6OCB-doped, 12 cholesterol-doped); the others were limited by localization noise, by undersampled slow modes or by non-thermal motion. The exponent $\alpha$ was obtained from an unconstrained power-law fit of $\lp$ versus $\Rf$, and the prefactor $\lp/\Rf^{2}$ from a fit with the exponent fixed at 2 (both fits weighted by the inverse squared half-width of each filament's 95\% confidence interval in $\ln\lp$, floored at 2\%; the confidence intervals on $\alpha$ and on the prefactor are 95\% percentile intervals from 4000 bootstrap resamples of filaments); $\Kb$ then follows from eqn~(5) of the main text. The noise model, the joint fit, the acceptance criteria and the sensitivity of $\Kb$ to the noise treatment are detailed in Sections~S3.2--S3.4. Figures~\ref{fig:si-checks} and \ref{fig:si-lpR} show, for every sample, the radius distribution, the compensated persistence length, the free exponent, the recording temperature and $\lp$ versus $\Rf$.

\section{Derivation of the persistence length of a smectic-A filament}
\label{si:theory}

\subsection{Elastic energy of the filament}

We model the smectic-A filament as a stack of concentric cylindrical
layers between an inner core radius $R_c$ and the outer filament radius
$R_{\mathrm f}$, with the director $\mathbf{n}$ normal to the layers. The
curvature elastic energy is~\cite{deGennes1993}
\begin{equation}
  F_\text{E} \;=\; \frac{K}{2}\int H^2\, \text{d}V
        \;,
  \label{eq:FE-def}
\end{equation}
where $K$ is the bend modulus, and
$H \equiv (c_1+c_2)/2$ is the mean curvature of the smectic layers ($K$ can be related to the nematic splay coefficient $K_1 = K/4$ by noting that $H = \tfrac{1}{2}(\nabla\cdot\hat{\mathbf{n}})$).

For a straight cylindrical filament of length $L$ the layers are
concentric cylinders, $H = 1/(2r)$, and
\begin{equation}
  F_\text{E}(0) \;=\; \frac{K}{2}\int_{0}^{L}\int_{R_\text{C}}^{R_{\mathrm F}} \frac{1}{4r^{2}}\; 2\pi r \, \text{d}r \, \text{d}l
         \;=\; \frac{\pi K L}{4} \,\ln\!\frac{R_{\mathrm F}}{R_\text{C}}.
  \label{eq:FE-straight}
\end{equation}

\subsection{Extra energy of a weakly bent filament}

Now let the filament be bent with a constant radius of curvature
$R_B = C^{-1}$, so that the layers become concentric segments of
tori. We use torus coordinates $(r,\nu,\psi)$: $r \in [R_c,R_{\mathrm f}]$ is
the distance from the filament centerline, $\nu \in [0,2\pi)$ is the
angle around the tube cross-section, and $\psi \in [0,L/R_B]$ is
the angle along the bent centerline (arc length $s = R_B\psi$, total
bend angle $\psi_{\max} = L/R_B$). The volume element is
\begin{equation}
  dV \;=\; r\,(R_B + r\cos\nu)\, d\nu\, d\psi\, dr .
  \label{eq:dV}
\end{equation}
Importantly, $\int dV = \pi L\ (R_{\mathrm f}^2 - R_c^2)$, is independent of $R_B$, such that
bending at fixed centerline length conserves the filament volume, so
$F_E(C)$ and $F_E(0)$ describe the same amount of material.

The layer at tube radius $r$ is a torus with principal curvatures
$c_1 = 1/r$ and $c_2 = \cos\nu/(R_B + r\cos\nu)$, so its mean
curvature is
\begin{equation}
  H(r,\nu) \;=\; \frac{R_B + 2r\cos\nu}{2r\,(R_B + r\cos\nu)} ,
  \qquad
  \nabla\cdot\mathbf{n} \;=\; 2H
  \;=\; \frac{R_B + 2r\cos\nu}{r\,(R_B + r\cos\nu)} .
  \label{eq:Hg}
\end{equation}
The elastic energy of the bent filament is therefore
\begin{equation}
  F_E(C) \;=\; \frac{K}{2}
  \int_{R_c}^{R_{\mathrm f}}\!\!\int_{0}^{L/R_B}\!\!\int_{0}^{2\pi}
  H^2 \;\, r\,(R_B + r\cos\nu)\; d\nu\, d\psi\, dr .
  \label{eq:FE-bent}
\end{equation}

\noindent For weak bending we introduce the small parameter
$\varepsilon \equiv r/R_B \ll 1$. The integrand becomes
\begin{equation}
  H^2\, dV
  \;=\; \frac{(R_B + 2r\cos\nu)^2}{4r\,(R_B + r\cos\nu)}\; d\nu\, d\psi\, dr
  \;=\; \frac{R_B}{4r}\,
        \frac{(1 + 2\varepsilon\cos\nu)^2}{1 + \varepsilon\cos\nu}\;
        d\nu\, d\psi\, dr .
  \label{eq:integrand}
\end{equation}
Expanding the fraction in powers of $\varepsilon$,
\begin{equation}
  \frac{(1 + 2\varepsilon\cos\nu)^2}{1 + \varepsilon\cos\nu}
  \;=\; 1 \;+\; 3\varepsilon\cos\nu \;+\; \varepsilon^{2}\cos^{2}\nu
        \;-\; \varepsilon^{3}\cos^{3}\nu \;+\; O(\varepsilon^{4}),
  \label{eq:expansion}
\end{equation}
so that
\begin{equation}
  H^2\, dV \;=\; \frac{R_B}{4r}
  \Big[\, 1 + 3\varepsilon\cos\nu + \varepsilon^{2}\cos^{2}\nu + \cdots \Big]
  \, d\nu\, d\psi\, dr .
\end{equation}
Integrating over $\nu$, the linear term vanishes,
$\int_0^{2\pi}\cos^2\nu\, d\nu = \pi$, and the cubic term also vanishes
($\int_0^{2\pi}\cos^3\nu\, d\nu = 0$), so the first neglected
contribution is $O(\varepsilon^4)$:
\begin{equation}
  \int_{0}^{2\pi} H^2\, dV
  \;=\; \left[\frac{\pi R_B}{2r} \;+\; \frac{\pi r}{4 R_B}\right]
  d\psi\, dr \;+\; O\!\big(\varepsilon^{4}\big) .
\end{equation}
Integrating over $\psi$ from $0$ to $L/R_B$,
\begin{equation}
  \int_{0}^{L/R_B}\!\!\int_{0}^{2\pi} H^2\, dV
  \;=\; \left[\frac{\pi L}{2r} \;+\; \frac{\pi L\, r}{4 R_B^{2}}\right] dr
  \;+\; O\!\big(\varepsilon^{4}\big),
\end{equation}
and finally over $r$ from $R_c$ to $R_{\mathrm f}$, with $C = 1/R_B$:
\begin{equation}
  \int H^2\, dV
  \;=\; \frac{\pi L}{2}\,\ln\!\frac{R_{\mathrm f}}{R_c}
  \;+\; \frac{\pi L}{8}\,\big(R_{\mathrm f}^{2} - R_c^{2}\big)\,C^{2}
  \;+\; O\!\big((C R_{\mathrm f})^{4}\big).
  \label{eq:Hg2-integral}
\end{equation}
Multiplying by $K/2$ as in Eq.~\eqref{eq:FE-def},
\begin{equation}
  F_E(C)
  \;=\; \frac{\pi K L}{4} \,\ln\!\frac{R_{\mathrm f}}{R_c}
  \;+\; \frac{\pi K L}{16}\,\big(R_{\mathrm f}^{2} - R_c^{2}\big)\,C^{2}
  \;+\; O\!\big((C R_{\mathrm f})^{4}\big)
  \;=\; F_E(0)
  \;+\; \frac{\pi}{16}\, K L\, R_{\mathrm f}^{2}\, C^{2}
  \;+\; \cdots,
  \label{eq:FE-final}
\end{equation}
where in the last step we used $R_{\mathrm f}^2 - R_c^2 \simeq R_{\mathrm f}^2$ for
$R_c \ll R_{\mathrm f}$. The extra elastic energy caused by bending the filament
is therefore
\begin{equation}
  \Delta F_{\mathrm{bend}}(C)
  \;=\; F_E(C) - F_E(0)
  \;\simeq\; \frac{\pi}{16}\, K L\, R_{\mathrm f}^{2}\, C^{2}.
  \label{eq:dF-bend}
\end{equation}

\subsection{Persistence length}

Comparing Eq.~\eqref{eq:dF-bend} with the worm-like-chain bending
energy~\cite{Kratky1949} of a filament of length $L$ at uniform centerline curvature $C$,
\begin{equation}
  \Delta F_{\mathrm{bend}} \;=\; \tfrac{1}{2}\,\kappa_f\, L\, C^{2},
  \qquad
  \kappa_f \;\equiv\; k_B T\, \lp,
\end{equation}
where $\kappa_f$ is the filament bending rigidity, we identify
\begin{equation}
  \kappa_f \;=\; \frac{\pi}{8}\, K R_{\mathrm f}^{2},
  \qquad
  \boxed{\;
  \lp \;=\; \frac{\pi}{8}\, \frac{K}{k_B T}\; R_{\mathrm f}^{2}
  \;}
  \label{eq:lp}
\end{equation}
which grows quadratically with the filament radius $R_{\mathrm f}$.

\section{Persistence length measurement from Fourier analysis}
\label{si:analysis}
\label{si:fourier}

\subsection{Filament tracking and Fourier decomposition}

Following Gittes \emph{et al.}~\cite{gittes1993flexural}, the shape of the filament
in each frame was characterized by its tangent angle $\theta(s,t)$, measured from the mean orientation of the contour, which
was decomposed into a superposition of cosine modes,
\begin{equation}
  \theta(s,t) \;=\; \sqrt{\tfrac{2}{L}}\,
  \sum_{n=1}^{\infty} a_n(t)\,
  \cos\!\Bigl(\frac{n\pi s}{L}\Bigr) ,
  \label{eq:si-decomposition}
\end{equation}
with mode amplitudes given by the Fourier inverse
\begin{equation}
  a_n(t) \;=\; \sqrt{\tfrac{2}{L}} \int_0^L \! \theta(s,t)\,
  \cos\!\Bigl(\frac{n\pi s}{L}\Bigr)\, \mathrm{d}s ,
  \qquad n = 1,2,\dots
  \label{eq:si-modes}
\end{equation}
Measuring $\theta$ from the mean orientation removes the zero-order ($n = 0$) term, which carries no bending energy. In practice the tracked centerline
of each frame was resampled into $N$ segments of length $\Delta s_k$ and
tangent angle $\theta_k$, and Eq.~\eqref{eq:si-modes} was evaluated as the
discrete sum $a_n \simeq \sqrt{2/L}\,\sum_{k=1}^{N} \theta_k\, \Delta s_k
\cos(n\pi s_k^{\mathrm{mid}}/L)$, where $s_k^{\mathrm{mid}}$ is the arc-length
midpoint of the $k$th segment.

For small bending the elastic energy is a quadratic sum over the modes,
$U = \tfrac{1}{2}\,\kappa \sum_{n\ge 1} (n\pi/L)^2 (a_n - a_n^0)^2$, where $\kappa$
is the bending rigidity and $a_n^0$ is the amplitude of the static
(time-averaged) shape, nonzero if the filament carries intrinsic
bends~\cite{gittes1993flexural}. Assuming the fluctuations are thermally driven, the equipartition theorem assigns $\tfrac{1}{2} k_B T$ to each mode independently, so that for
$n \ge 1$
\begin{equation}
  \operatorname{Var}(a_n) \;=\;
  \bigl\langle (a_n - a_n^0)^2 \bigr\rangle \;=\;
  \frac{k_B T}{\kappa}\Bigl(\frac{L}{n\pi}\Bigr)^{\!2} \;=\;
  \frac{L^{2}}{\lp\, \pi^{2} n^{2}},
  \label{eq:si-equipartition}
\end{equation}
with $\lp = \kappa/k_B T$ the persistence length. Because the static shape
$a_n^0$ is not known a priori, the variance of each $a_n(t)$ was
computed about its time average, which removes any static (intrinsic-bend)
contribution. The absence of cross-terms in $U$ implies that the amplitudes
of different modes fluctuate independently, so each mode provides an
independent estimate of $\lp$.

\subsection{Noise model and joint fit}

Localization error adds a bias, not merely a random error, to every measured
variance~\cite{gittes1993flexural}. Each of the $N+1$ sampled points deviates from
the true filament position by a random distance $\varepsilon_k$; for
independent, identically distributed errors of variance
$\sigma^{2} \equiv \langle \varepsilon_k^{2} \rangle$, the contribution to
the variance of each mode is~\cite{gittes1993flexural,Brangwynne2007}
\begin{equation}
  \operatorname{Var}_{\mathrm{noise}}(a_n) \;=\;
  \bigl\langle (a_n^{\mathrm{noise}})^{2} \bigr\rangle \;=\;
  \frac{4\sigma^{2}}{L}
  \left[\, 1 + (N-1)\,
  \sin^{2}\!\Bigl(\frac{n\pi}{2N}\Bigr) \right],
  \label{eq:si-noise}
\end{equation}
which is flat at low $n$ and rises toward high $n$, opposite to the thermal
$n^{-2}$ decay of Eq.~\eqref{eq:si-equipartition}. The measured spectrum was
therefore fit, from the first mode of the accepted window (see \emph{Acceptance criterion} below) up to $n = 12$, to the two-parameter
model~\cite{gittes1993flexural}
\begin{equation}
  \operatorname{Var}_{\mathrm{meas}}(a_n) \;=\;
  \frac{1}{\lp}\Bigl(\frac{L}{n\pi}\Bigr)^{\!2}
  \;+\;
  \frac{4\sigma^{2}}{L}
  \left[\, 1 + (N-1)\,
  \sin^{2}\!\Bigl(\frac{n\pi}{2N}\Bigr) \right],
  \label{eq:si-jointfit}
\end{equation}
with $\lp$ and $\sigma$ as the only free parameters, fitted together by least
squares on $\ln\operatorname{Var}(a_n)$; the thermal exponent is held fixed at
$-2$. The modes above the accepted window, where the noise term becomes
comparable to the thermal term, fix $\sigma$. Tracking error on our centerlines
is correlated along the filament rather than independent from point to point
as assumed in Eq.~\eqref{eq:si-noise}, so the fitted $\sigma$ is an effective
noise level and not a localization precision; Section~S3.4 gives the
sensitivity of $\lp$ to this choice. Statistical uncertainty on $\lp$ was obtained by a
bootstrap: frames were resampled with replacement 300 times, the mode
variances and the joint fit recomputed for each resample, and the 2.5th and
97.5th percentiles of the resulting $\lp$ distribution quoted as the 95\,\%
confidence interval.

In summary, the reported $\lp$ of each filament is obtained in four steps:
(i) a window of thermal modes is selected and the filament is accepted as
described in Section~S3.3; (ii) Eq.~\eqref{eq:si-jointfit} is fitted to
$\ln\operatorname{Var}(a_n)$ from the first mode of that window up to $n = 12$,
with $\lp$ and $\sigma$ free (the fit starts at mode 1 for 42 of the 89 accepted
filaments and at modes 2--7 for the others); (iii) its 95\,\% confidence
interval comes from 300 bootstrap resamples of the frames; (iv) $\Kb$ of the
filament follows from eqn~(5) of the main text, and the $\Kb$ of a sample is
the mean over its filaments (Fig.~2d of the main text) or the weighted slope-2
prefactor of $\lp$ versus $\Rf$.

\subsection{Acceptance criterion}

A filament measurement was accepted only if its variance spectrum contains a
mode window that is quantitatively consistent with thermal bending. All
contiguous windows of 3--7 modes within modes 1--12 were scanned, and a
window qualifies only if it satisfies \emph{all} of the following
pre-declared conditions:
\begin{enumerate}
  \item \textbf{Thermal slope:} the log--log slope of
        $\operatorname{Var}(a_n)$ vs.\ $n$ is $-2 \pm 0.45$, with
        goodness of fit $R^{2} \ge 0.90$;
  \item \textbf{Plateau consistency:} the per-mode estimates
        $\lp(n) = L^{2}/\pi^{2} n^{2} \operatorname{Var}(a_n)$
        are constant across the window, with robust
        coefficient of variation $\le 0.30$;
  \item \textbf{Mode independence:} the maximum absolute Pearson correlation
        between the amplitudes of any two window modes is $\le 0.55$, as
        required by equipartition for independent modes.
\end{enumerate}
If at least one window passes, the widest passing window is used and a joint
fit of Eq.~\eqref{eq:si-jointfit} is performed on it; this fit must reach
$R^{2} \ge 0.8$ with a noise level $\sigma \le 250$~nm. The first mode of the
window then starts the reported fit of Section~S3.2. A measurement is accepted
only if it also meets the following conditions:
\begin{enumerate}
\item \textbf{Sampling:} every window mode is recorded for at least five of its
      own relaxation times $\tau_n = \tau_1/n^{4}$, with $\tau_1$ from an
      exponential fit to the autocorrelation of $a_1(t)$. If not, the filament
      is refitted from the first well-sampled mode, on the window whose slope
      is closest to $-2$;
\item \textbf{Low modes:} modes 1 and 2 exceed the window plateau by at most
      60\,\%. If not, the lowest mode of the window is raised by one, up to
      mode 4;
\item \textbf{Stable shape:} the 5--95\,\% spread of the chord deviation is at
      most 5\,\% of $L$, the smoothed centroid moves by at most 10~$\mu$m within
      25 frames, and the smoothed length changes by at most 60\,\% during the
      movie;
\item \textbf{Size and precision:} $\Rf \le 2.3~\mu$m, and the half-width of the
      95\,\% confidence interval of $\Kb$ is at most 25\,\% of $\Kb$, for both
      the window fit and the reported fit;
\item \textbf{Inspection:} every filament passing these tests was inspected on
      the wide-field images, and traces that were segments of longer filaments
      or arcs of loops, or that had an attached end, were excluded.
\end{enumerate}
Measurements failing the criterion are dominated by localization noise, by
undersampled slow modes, or by non-thermal motion, and carry no usable $\lp$.

\clearpage   

\end{document}